\documentclass{article}

\usepackage{arxiv}

\usepackage[utf8]{inputenc} 
\usepackage[T1]{fontenc}    
\usepackage{hyperref}       
\usepackage{url}            
\usepackage{booktabs}       
\usepackage{amsfonts}       
\usepackage{nicefrac}       
\usepackage{microtype}      
\usepackage{lipsum}		
\usepackage{graphicx}
\usepackage{multirow}
\usepackage{cite}

\usepackage{amsmath}
\usepackage{chemformula}

\usepackage{epstopdf}
\usepackage{array}
\usepackage{subfigure}
\usepackage{ragged2e}
\usepackage{booktabs,makecell,multirow,tabularx}
\usepackage[ruled,linesnumbered]{algorithm2e}

\title{DeepPropNet: an operator learning-based predictor for thermal plasma properties}

\author{Zuo Wang, ~Linlin Zhong\thanks{This paper is currently under consideration by a journal.} \\
	School of Electrical Engineering\\
	Southeast University\\
	No 2 Sipailou, Nanjing, Jiangsu Province 210096, China\\
	\texttt{linlin@seu.edu.cn}\\
}

\date{April 27, 2026}

\renewcommand{\headeright}{}
\renewcommand{\undertitle}{A Preprint}

\begin{document}
\maketitle

\begin{abstract}
	Thermal plasma properties play a critical role in plasma simulations and plasma-related applications. However, their strong nonlinear dependence on temperature, pressure, and gas composition makes accurate and efficient evaluation challenging. In this work, an operator learning-based model, termed DeepPropNet, is proposed for fast prediction of thermodynamic and transport properties of thermal plasmas. Two architectures are developed, including a single-property model (S-DeepPropNet) and a Mixture of Experts (MoE)-based multi-properties model (MoE-DeepPropNet). The proposed models learn the nonlinear mapping from plasma operating conditions to physical properties based on high-fidelity datasets. The MoE architecture enables efficient multi-property prediction within a unified framework. Predictions are performed for binary SF$_6$-N$_2$ and ternary C$_4$F$_7$N-CO$_2$-O$_2$ mixtures. The results show that the proposed models achieve high accuracy, with relative $L^2$ errors on the order of $10^{-3}$ to $10^{-2}$, while maintaining strong generalization capability under unseen conditions. The applicability of DeepPropNet is further demonstrated by coupling with finite volume method (FVM) and physics-informed neural networks (PINNs). The results indicate that DeepPropNet provides an efficient and scalable approach for plasma property prediction and plasma simulations.
\end{abstract}

\section{Introduction}
\label{sec:sec1}
\paragraph{}
Thermal plasmas, characterized by highly ionized gases at elevated temperatures, play a crucial role in various industrial and scientific applications, such as arc welding \cite{cite1}, circuit breakers \cite{cite2}, plasma torches \cite{cite3}, and aerospace engineering \cite{cite4}. Accurate knowledge of thermal plasma properties, particularly thermodynamic properties and transport coefficients, is essential for optimizing performance and improving numerical simulations of plasma-related processes. Traditionally, these properties are obtained through theoretical calculations and numerical models based on equilibrium assumptions, solving complex equations derived from statistical mechanics and kinetic theory. For example, plasma transport coefficients including electrical conductivity, thermal conductivity, and viscosity, are usually numerically determined by the Chapman-Enskog method, which requires calculating various complex transport cross sections and multiple collision integrals \cite{cite5}. These methods are often computationally expensive, making real-time applications and high-fidelity simulations challenging.

\paragraph{}
Moreover, the thermophysical properties of thermal plasmas are highly dependent on gas species, their mixing ratios, and local thermodynamic state variables (e.g., temperature and pressure). For example, sulfur hexafluoride (SF$_6$), a widely used insulating and arc quenching gas in power industry, is usually mixed with buffer gases such as N$_2$ and CO$_2$, to fulfill the special industrial application requirements under low-temperature or high-pressure operation conditions \cite{cite6}. Consequently, the thermal plasma properties need to be calculated under different gas compositions, temperatures, and pressures to meet the needs of plasma fluid simulation. In practical applications, to reduce the computational cost, plasma properties are typically precomputed under selected conditions and then accessed via table-lookup and interpolation during simulations. However, this suffers from high storage cost and reduced accuracy in high-dimensional and strongly nonlinear regimes, especially for extrapolation beyond the tabulated range.

\paragraph{}
In recent years, deep learning has emerged as a powerful tool for approximating complex physical relationships with remarkable efficiency \cite{cite7, cite8}. Leveraging neural networks to learn plasma property mappings from precomputed datasets could provide a fast and accurate alternative to traditional numerical approaches. While conventional deep learning models map between finite-dimensional discrete vectors, operator learning has demonstrated strong generalization capability by approximating nonlinear operators that map between infinite-dimensional function spaces (i.e., Banach spaces) \cite{cite9, cite10, cite11}. Architectures such as Fourier neural operator (FNO) \cite{cite12} and deep operator network (DeepONet) \cite{cite13} have been proposed and applied to solve various problems in scientific computation, such as fluid dynamics \cite{cite14}, chemical kinetics \cite{cite15}, and multiscale simulations \cite{cite16}. In plasma physics, operator learning has been used as a surrogate model to predict plasma dynamics. For example, Gopakumar et al. \cite{cite17} developed an FNO-based surrogate model to predict plasma evolution within a Tokamak reactor, while Bonotto et al. \cite{cite18} proposed a convolutional physics-informed neural operator for plasma equilibrium reconstruction and separatrix reconstruction. However, few studies have applied operator learning to model the complex nonlinear mapping from plasma operating conditions to plasma properties, highlighting the need for a unified and efficient framework to capture such relationships.

\paragraph{}
In this work, we propose DeepPropNet, an operator learning-based predictive model designed to rapidly estimate a wide range of essential thermal plasma properties, including thermodynamic properties (i.e., mass density, enthalpy, entropy, and specific heat) and transport coefficients (i.e., electrical conductivity, viscosity, and thermal conductivity). DeepPropNet has two specific architectures, namely S-DeepPropNet and MoE-DeepPropNet. The former is designed for single-property prediction using a single network, and the latter is designed for multi-properties prediction based on Mixture of Experts (MoE) structure \cite{cite19, cite20}. Both S-DeepPropNet and MoE-DeepPropNet are trained on high-fidelity datasets obtained from conventional theoretical calculations and aim to deliver real-time predictions with minimal computational overhead while maintaining high accuracy.

\paragraph{}
The rest of this paper is organized as follows. Sec. \ref{sec:sec2} presents the methodology behind DeepPropNet, including data preparation, neural network architecture, and training strategies. Sec. \ref{sec:sec3} provides validation results and performance comparisons against traditional methods, covering the prediction of thermal plasma properties of binary gases (i.e., SF$_6$-N$_2$) and ternary gases (i.e., C$_4$F$_7$N-CO$_2$-O$_2$). We also demonstrate the thermal plasma simulation by combining DeepPropNet with finite volume method (FVM) and physics-informed neural network (PINN), respectively. Finally, Sec. \ref{sec:sec4} concludes the study with potential applications and future research directions.

\section{Methodology}
\label{sec:sec2}
\paragraph{}
Thermal plasma properties considered in this work include both thermodynamic and transport properties. Their evaluation is based on equilibrium composition calculations, followed by the determination of thermodynamic properties and transport coefficients using kinetic theory. However, these conventional approaches involve strong multi-physics coupling and high computational cost, especially for multi-component plasmas over wide operating conditions. To address this limitation, an operator learning-based surrogate model, termed DeepPropNet, is developed to directly map plasma operating conditions to the corresponding properties. Two architectures are constructed: the S-DeepPropNet for individual property prediction and the MoE-DeepPropNet for simultaneous multi-properties prediction. The physical formulations and model architectures are described in the following subsections.

\subsection{Thermodynamic and transport properties of thermal plasmas}
\label{sec:sec2.1}
\paragraph{}
Since our previous studies \cite{cite21, cite22, cite23} have explored the detailed calculation procedures for thermodynamic and transport properties of thermal plasmas, only a brief overview is presented here. The calculation of thermal plasma properties requires the knowledge of plasma compositions, which are usually determined by the minimization of Gibbs free energy of a plasma system. The Gibbs free energy $G$ of a plasma system with $N$ species can be expressed as \cite{cite24}

\begin{equation}
	\label{equ:equ1}
	G = \sum_{i=1}^{N} n_i \mu_i
\end{equation}

where $n_i$ and $\mu_i$ are the number density and chemical potential of species $i$, respectively. The latter is usually derived from species’ partition functions. According to thermodynamic theory, any spontaneous process in a plasma system at constant temperature and pressure will decrease the Gibbs free energy of the system until reaching equilibrium. This means we can obtain the plasma composition by searching for the minimum Gibbs free energy, where $dG = 0$.

\paragraph{}
Once the equilibrium compositions are determined, the thermodynamic properties of the plasma mixture can be calculated according to their fundamental definitions~\cite{cite21}. Four thermodynamic properties are evaluated: mass density ($\rho$), specific enthalpy ($h$), specific entropy ($s$), and specific heat at constant pressure ($C_p$). The mass density is given by the sum of the products of species number densities and their respective particle masses. Furthermore, the specific enthalpy and entropy are derived from the standard-state molar enthalpy and entropy related to the partition functions, whereas $C_p$ is determined by the temperature derivative of the enthalpy. Moreover, high-temperature thermal plasmas deviate significantly from the ideal gas law due to the long-range Coulomb interactions among the dense population of electrons and ions. To accurately capture the real gas effects, the Debye--Hückel (DH) corrections should be incorporated.

\paragraph{}
Following the determination of thermodynamic properties, the transport coefficients of the thermal plasma, namely electrical conductivity ($\sigma$), thermal conductivity ($\kappa$), and viscosity ($\eta$), are conventionally evaluated by solving the Boltzmann equation based on the Chapman-Enskog approximation~\cite{cite22}. The accuracy of this evaluation is fundamentally governed by the collision integrals, which encapsulate the microscopic collision dynamics for all possible interacting pairs of species $i$ and $j$ in the mixture. The fundamental expression for the temperature-dependent collision integral characterized by the order $(l, s)$ is given by:

\begin{equation}
\label{equ:equ2}
\Omega_{ij}^{(l,s)} =
\frac{4(l+1)}{\pi (s+1)! \left[ 2l + 1 - (-1)^l \right]}
\int_{0}^{\infty}
e^{-\gamma_{ij}^2} \gamma_{ij}^{2s+3}
Q_{ij}^{(l)}(\gamma_{ij}) \, d\gamma_{ij}
\end{equation}

where $k_B$ is the Boltzmann constant, $T$ is the temperature, $\mu_{ij}$ is the reduced mass of the colliding pair, and $\gamma$ is the reduced relative velocity. $Q_{ij}^{(l)}$ represents the transport cross-section and is obtained by integrating the energy-dependent differential cross section over the scattering deflection angle~\cite{cite25}. Furthermore, these integrations strictly depend on diverse interparticle interaction potentials, requiring distinct theoretical treatments for four types of interactions: neutral-neutral, neutral-ion, electron-neutral, and charged-charged interactions~\cite{cite23}. As the complexity of the plasma mixture increases, the number of interacting pairs scales quadratically.

\paragraph{}
Consequently, determining the transport coefficients requires the numerical evaluation of multiple integrals for each particle pair, tightly coupled with the high-order matrix inversions required by the Chapman-Enskog formulation. These characteristics highlight that the evaluation of thermal plasma properties involves strong nonlinearity and complex coupling across temperature, pressure, and composition, which poses significant challenges for efficient numerical evaluation and motivates the development of data-driven surrogate models.

\subsection{Single DeepPropNet (S-DeepPropNet) for single-property prediction}
\label{sec:sec2.2}
\begin{figure}
	\centering
	\includegraphics[width=10cm]{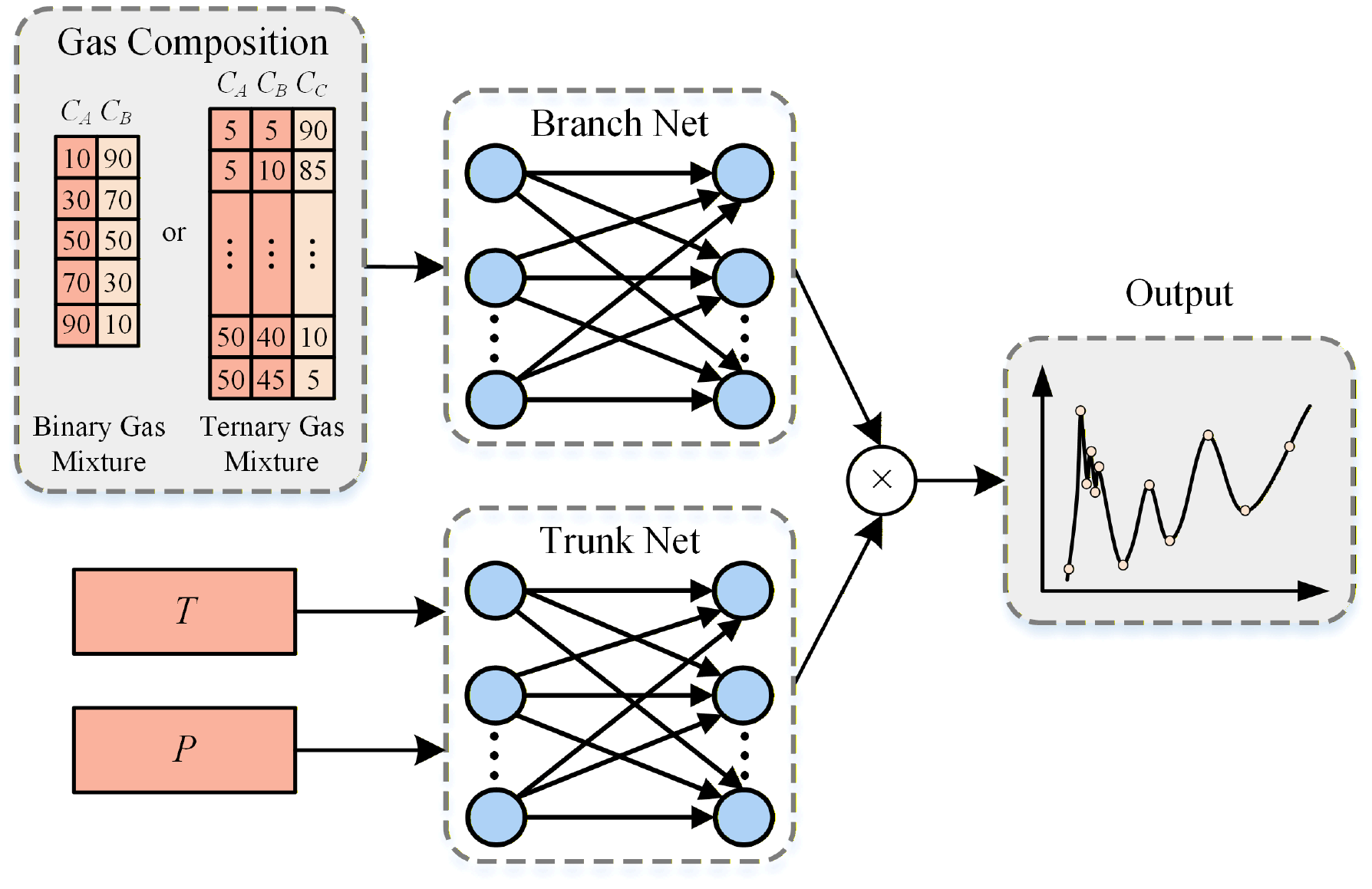}
	\caption{Diagram of S-DeepPropNet for single-property prediction.}
	\label{fig:fig1}
\end{figure}

\paragraph{}
The Deep Operator Network (DeepONet), originally proposed by Lu et al. \cite{cite13}, provides a framework for learning nonlinear mappings between function spaces based on the universal approximation theorem for operators. Its architecture consists of two components: a branch network that encodes the input function into a finite-dimensional latent representation and a trunk network that evaluates the spatial coordinates to construct the corresponding basis functions. Building upon this framework, the S-DeepPropNet is developed, as illustrated in Fig.\ref{fig:fig1}, to predict individual thermal plasma properties under varying operating conditions.

\paragraph{}
Specifically, the branch network takes the operational condition vector $\mathbf{C}$, representing the gas mixing proportions, as input and encodes it into a latent representation $\mathbf{v} = [v_1, v_2, \ldots, v_p]$ through fully connected layers. Concurrently, the trunk network takes the query vector $\mathbf{y}$ as input. For thermodynamic and transport properties, $\mathbf{y}$ consists of the temperature $T$ and the pressure $P$. For radiation property prediction, the model can be extended by augmenting the trunk input with an additional dimension. Specifically, the plasma radius $R$ is included as an extra input variable~\cite{cite26}. The trunk network then outputs a corresponding set of basis functions $\mathbf{u} = [u_1, u_2, \ldots, u_p]$ through its own fully connected layers.

\paragraph{}
The predicted property is obtained as the inner product of the outputs from the branch and trunk networks, followed by the addition of a global bias. The forward mapping can be expressed as

\begin{equation}
\label{equ:equ3}
G_{\theta}(\mathbf{C})(y) = \sum_{k=1}^{p} b_k(\mathbf{C}) \, t_k(y) + b_0
\end{equation}

where $\theta$ denotes all trainable parameters, $p$ is the latent dimension, $b_0$ is a global bias, and $b_k$ and $t_k$ are the outputs of the branch network and trunk network, respectively. The model is trained by minimizing the mean squared error (MSE) between the predicted and reference values over the sampled input pairs $(\mathbf{C}, y)$.

\subsection{Mixture of Experts (MoE)-based DeepPropNet (MoE-DeepPropNet) for multi-properties prediction}
\label{sec:sec2.3}
\begin{figure}[!b]
	\centering
	\includegraphics[width=9cm]{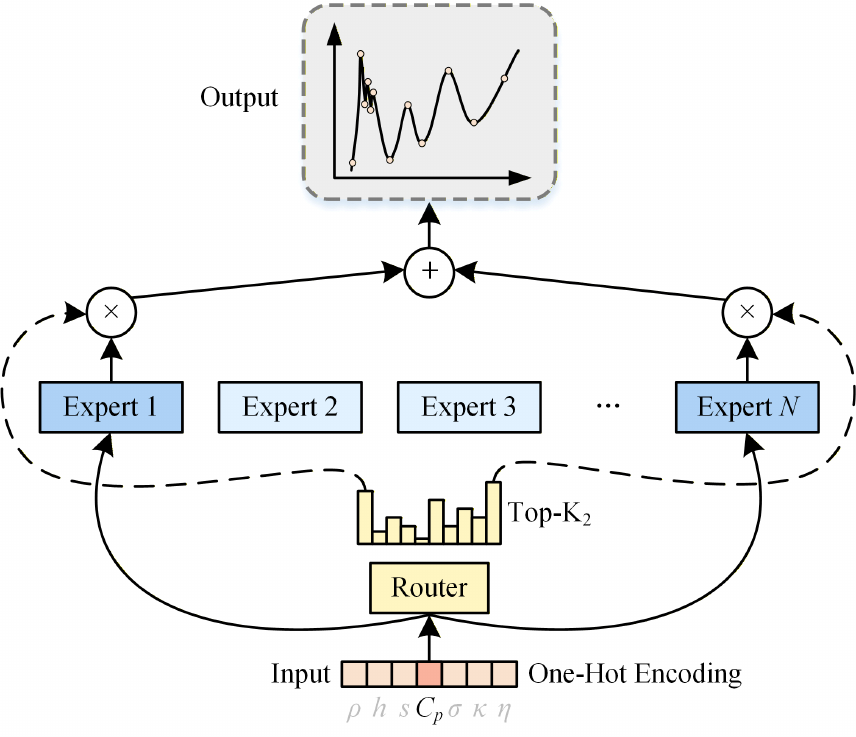}
	\caption{Diagram of MoE-DeepPropNet for multi-property prediction.}
	\label{fig:fig2}
\end{figure}

\paragraph{}
Since S-DeepPropNet is formulated for single-property prediction, a straightforward extension to multiple plasma properties would require separate models for each target, resulting in significant computational redundancy. Moreover, although different properties share common input variables $(\mathbf{C}, P, T)$, their underlying dependencies on these variables can vary substantially. To address these challenges, a Mixture of Experts (MoE)-based architecture, termed MoE-DeepPropNet, is developed, as illustrated in Fig.~\ref{fig:fig2}.

\paragraph{}
By consolidating multiple prediction tasks into a single framework, the proposed model enables efficient multi-property evaluation within a unified forward pass. In this architecture, the prediction of the target property vector $\mathbf{Y}$ is formulated as a weighted summation of multiple expert outputs.

\begin{equation}
\label{equ:equ4}
\mathbf{Y}(I, \mathbf{C}, y) = \sum_{i=1}^{N} G_i(I)\, E_i(\mathbf{C}, y)
\end{equation}

where $N$ is the total number of expert subnetworks, $E_i(\mathbf{C}, y)$ denotes the output of the $i$-th expert, and $G_i(I)$ represents the sparse routing weight assigned by the gating network. The vector $I$ is a one-hot encoded indicator specifying the target plasma property.

\paragraph{}
The expert subnetworks $E_i$ inherit the foundational branch and trunk architecture described in Sec.\ref{sec:sec2.2}. To balance computational cost and expert utilization, a sparsely gated noisy Top-$K$ strategy is adopted. This mechanism employs a gating network together with a noise network to map the one-hot encoded identifier $I$ to routing weights. The gating and noise networks output the clean logits $h(I)$ and noise parameters $\nu(I)$, respectively. To promote load balancing and prevent expert over-specialization, Gaussian noise is injected into the gating logits during training. The noisy logit for the $i$-th expert is given by

\begin{equation}
\label{equ:equ5}
\tilde{h}_i(I) = h_i(I) + \epsilon \cdot \mathrm{Softplus}\!\big(\nu_i(I)\big)
\end{equation}

where $\epsilon \sim \mathcal{N}(0, 1)$ is sampled from a standard normal distribution.

\paragraph{}
Subsequently, a selection operator identifies the index set $S_K$ corresponding to the top-$K$ experts with the largest noisy logits. To enforce sparsity, the logits of unselected experts are masked with negative infinity ($-\infty$). The routing weight $G_i(I)$ in Eq.~\eqref{equ:equ4} is then computed by applying the Softmax function over the masked logits.

\begin{equation}
\label{equ:equ6}
G_i(I) = \frac{\exp\big(\tilde{h}_i(I)\big)}{\sum_{j=1}^{N} \exp\big(\tilde{h}_j(I)\big)}
\end{equation}

where the masked logit $\hat{h}_i(I)$ is defined as

\begin{equation}
\label{equ:equ7}
\hat{h}_i(I) =
\begin{cases}
	\tilde{h}_i(I), & \text{if } i \in S_K, \\
	-\infty, & \text{otherwise}
\end{cases}
\end{equation}

\paragraph{}
Since $\exp(-\infty) = 0$, this masking ensures that unselected experts receive zero routing weight. By activating only a subset of experts in each forward pass, the computational cost is reduced, while the injected noise introduces stochasticity that facilitates load balancing during training. More importantly, it enables different experts to specialize in distinct mappings between $(\mathbf{C}, P, T)$ and target properties, thereby accommodating property-specific dependencies within a unified model.

\section{Numerical experiments}
\label{sec:sec3}
\paragraph{}
To evaluate the proposed operator learning-based frameworks, a series of numerical experiments are conducted on complex thermal plasmas. This section examines the performance of DeepPropNet across four scenarios. First, S-DeepPropNet and MoE-DeepPropNet are used to predict the thermodynamic and transport properties of binary SF$_6$-N$_2$ and ternary C$_4$F$_7$N-CO$_2$-O$_2$ mixtures. Next, the practical applicability of the proposed model is demonstrated by coupling it with finite volume method (FVM) for thermal plasma simulation and with physics-informed neural networks (PINNs). All training datasets are generated based on the theoretical formulations described in Sec.\ref{sec:sec2}. The neural networks are implemented in the open-source framework PyTorch~\cite{cite27} and trained using the Adam optimizer.

\subsection{Predicting plasma properties of binary gas mixtures}
\label{sec:sec3.1}
\paragraph{}
The thermodynamic and transport properties of binary SF$_6$-N$_2$ thermal plasma are considered to evaluate the predictive performance of both S-DeepPropNet and MoE-DeepPropNet. The training datasets are generated over a temperature range of 300--30000~K with a uniform sampling interval of 100~K. The SF$_6$ mixing proportions are set to 10\%, 30\%, 50\%, 70\%, and 90\%. For the operating pressure, S-DeepPropNet is trained using logarithmically spaced values from 1.0 to 15.85~bar (1.0, 1.58, 2.51, 3.98, 6.31, 10.0, and 15.85~bar), providing higher resolution in the low-pressure regime where plasma properties exhibit strong variations~\cite{cite21,cite22}. In contrast, MoE-DeepPropNet adopts linearly spaced pressures from 1 to 15~bar with an interval of 2~bar. Regarding network configurations, S-DeepPropNet consists of five hidden layers with 200 neurons per layer and is trained separately for each property over 100,000 epochs. MoE-DeepPropNet employs ten experts, each composed of four hidden layers with 100 neurons. The gating and noise networks each contain one hidden layer with 50 neurons, and a Top-2 sparse activation strategy is adopted. The model is trained for 1,000,000 epochs. Both models are optimized using the Adam optimizer with a constant learning rate of $10^{-4}$.

\begin{figure}[!b]
	\centering
	\includegraphics[width=15cm]{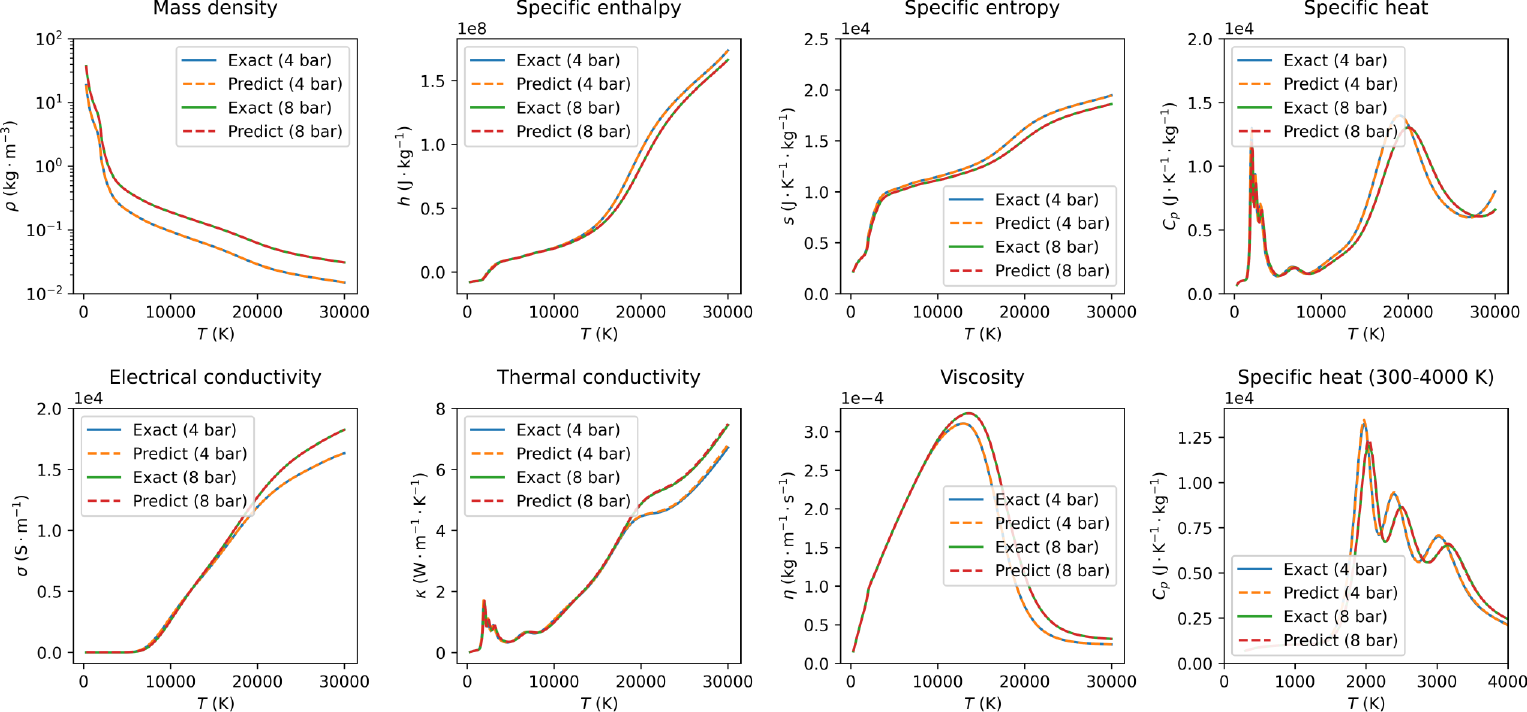}
	\caption{Prediction of thermodynamic and transport properties of SF$_6$-N$_2$ (75\%-25\%) plasmas under different pressures using MoE-DeepPropNet.}
	\label{fig:fig3}
\end{figure}

\paragraph{}
The predictive capability of MoE-DeepPropNet is first evaluated under unseen operating conditions. Fig.~\ref{fig:fig3} presents the predicted thermodynamic and transport properties of SF$_6$-N$_2$ (75\%-25\%) plasma mixture under different pressures. The 75\% SF$_6$ composition is intentionally excluded from the training set to assess the interpolation capability of the model. Overall, the predicted curves show excellent agreement with the reference results over the entire temperature range at both 4 and 8~bar, indicating that the model generalizes well to unseen mixture compositions.

\paragraph{}
More specifically, MoE-DeepPropNet accurately reproduces the major thermophysical trends of the plasma. The mass density decreases monotonically with temperature, while the specific enthalpy and entropy increase progressively as the plasma evolves from a molecular state to a highly ionized state. For the specific heat capacity $C_p$, the model captures both the low-temperature multi-peak structure and the broader high-temperature peak, which are associated with successive dissociation and ionization processes. A similarly good agreement is observed for the transport properties. In particular, the rapid increase in electrical conductivity $\sigma$ at high temperatures, the non-monotonic variation in thermal conductivity $\kappa$, and the characteristic peak in viscosity $\eta$ are all well reproduced. These results indicate that the proposed model can preserve the strong nonlinear responses of plasma properties to temperature under different pressure conditions.

\begin{figure}
	\centering
	\includegraphics[width=15cm]{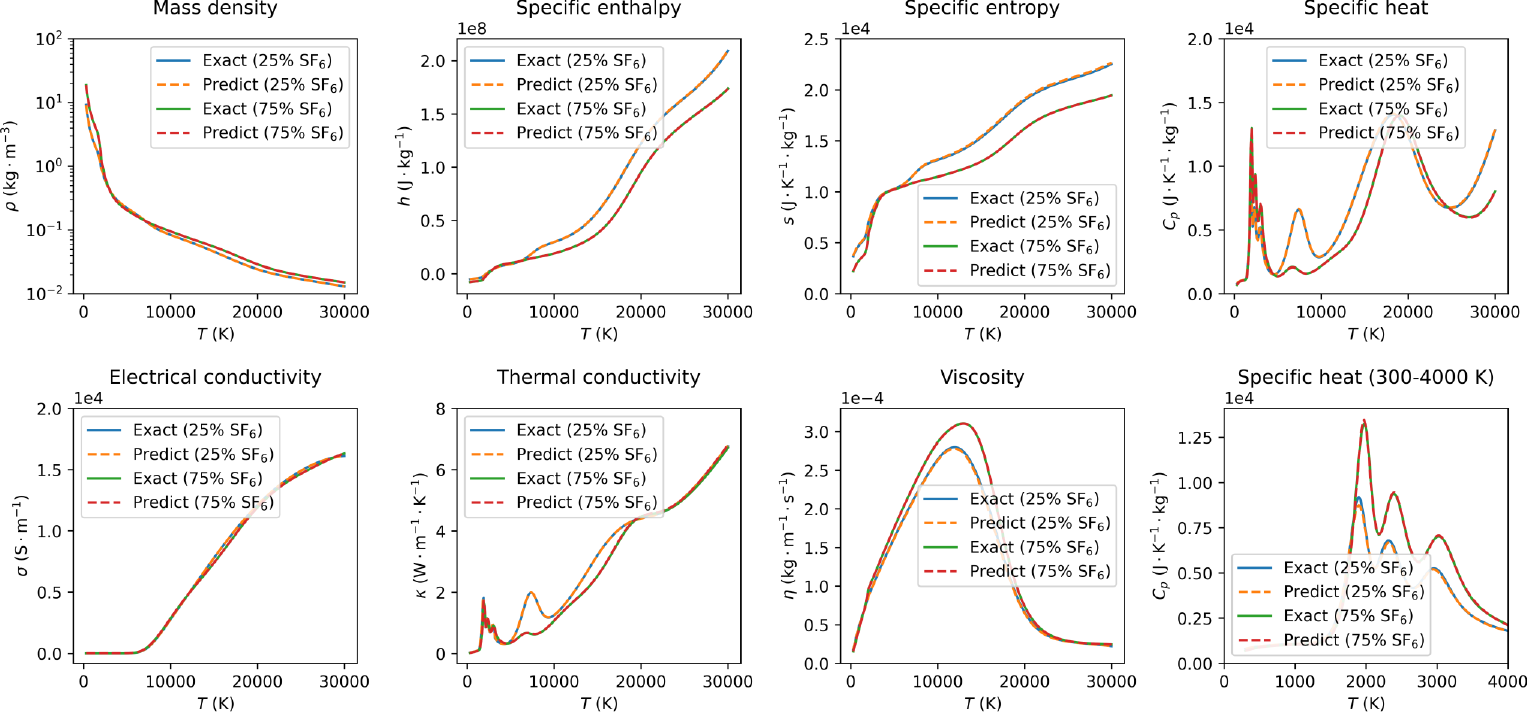}
	\caption{Prediction of thermodynamic and transport properties of SF$_6$-N$_2$ plasmas at 4 bar with different gas mixing proportions using MoE-DeepPropNet.}
	\label{fig:fig4}
\end{figure}

\paragraph{}
Fig.~\ref{fig:fig4} further examines the predictive performance under different gas compositions at a fixed pressure of 4~bar. The predicted results exhibit excellent agreement with the reference data over the entire temperature range for both 25\% and 75\% SF$_6$ mixtures, demonstrating the model’s ability to generalize across varying compositions. As the SF$_6$ fraction increases, systematic shifts in the thermodynamic and transport properties are clearly observed and accurately captured by the model. In particular, higher SF$_6$ content leads to increased mass density and shifts the onset of rapid electrical conductivity growth to higher temperatures, reflecting delayed ionization processes. These trends are well reproduced by MoE-DeepPropNet.

\paragraph{}
For the specific heat capacity $C_p$, the model accurately resolves the pronounced multi-peak structures in the low-temperature range (300-4000~K). These peaks are associated with successive molecular dissociation reactions and are highly sensitive to composition. Similarly, for the transport properties, the model reproduces the non-monotonic behavior of thermal conductivity $\kappa$ and the characteristic peak in viscosity $\eta$, both of which shift with changing composition. The close agreement in these features suggests that the model is able to represent the composition-dependent variations in collision processes and energy transport mechanisms. Overall, these results confirm that MoE-DeepPropNet can accurately capture the coupled effects of temperature and composition on plasma properties, even in regions with strong nonlinear variations.

\paragraph{}
To quantitatively assess the performance of the MoE-DeepPropNet against the S-DeepPropNet, the relative $L^2$ errors for all seven properties are evaluated at unseen test pressures (4, 8, 12, and 16~bar) and averaged over four SF$_6$ mixing ratios (20\%, 40\%, 60\%, and 80\%). The results are summarized in Table \ref{tab:tab1}. Both models achieve high predictive accuracy, with relative errors predominantly on the order of $10^{-3}$ across most operating conditions. At moderate pressures (4--12~bar), the two models exhibit comparable performance for most properties, with only minor differences in error levels. More pronounced differences emerge at the highest pressure (16~bar), where the plasma properties exhibit stronger nonlinear variations. In this regime, MoE-DeepPropNet provides improved accuracy for several properties. In particular, the relative error in the specific heat capacity $C_p$ is reduced from 0.0198 to 0.0035, while the error in thermal conductivity $\kappa$ decreases from 0.0113 to 0.0077. Similar improvements are observed for density $\rho$, entropy $s$, and viscosity $\eta$, indicating enhanced robustness of the MoE architecture under high-pressure conditions.

\begin{table}[htbp]
	\centering
	\caption{Relative $L^2$ errors of thermodynamic and transport properties for SF$_6$-N$_2$ plasmas predicted by S-DeepPropNet and MoE-DeepPropNet under different pressures, averaged over four gas mixing ratios.}
	\label{tab:tab1}
	\begin{tabular}{|c|c|c|c|c|c|c|c|c|}
		\hline
		Pressure (bar) & Model & $\rho$ & $h$ & $s$ & $C_p$ & $\sigma$ & $\kappa$ & $\eta$ \\
		\hline
		
		\multirow{2}{*}{4}
		& S-DeepPropNet  & 0.0040 & 0.0012 & 0.0015 & 0.0034 & 0.0010 & 0.0052 & 0.0028 \\
		& MoE-DeepPropNet& 0.0042 & 0.0013 & 0.0015 & 0.0034 & 0.0010 & 0.0050 & 0.0027 \\
		\hline
		
		\multirow{2}{*}{8}
		& S-DeepPropNet  & 0.0043 & 0.0013 & 0.0015 & 0.0031 & 0.0010 & 0.0048 & 0.0026 \\
		& MoE-DeepPropNet& 0.0045 & 0.0013 & 0.0014 & 0.0029 & 0.0010 & 0.0046 & 0.0026 \\
		\hline
		
		\multirow{2}{*}{12}
		& S-DeepPropNet  & 0.0047 & 0.0013 & 0.0014 & 0.0028 & 0.0010 & 0.0043 & 0.0025 \\
		& MoE-DeepPropNet& 0.0058 & 0.0014 & 0.0015 & 0.0035 & 0.0011 & 0.0044 & 0.0026 \\
		\hline
		
		\multirow{2}{*}{16}
		& S-DeepPropNet  & 0.0093 & 0.0032 & 0.0044 & 0.0198 & 0.0026 & 0.0113 & 0.0043 \\
		& MoE-DeepPropNet& 0.0040 & 0.0012 & 0.0012 & 0.0035 & 0.0009 & 0.0077 & 0.0020 \\
		\hline
		
		\multirow{2}{*}{Overall}
		& S-DeepPropNet  & 0.0076 & 0.0095 & 0.0018 & 0.0107 & 0.0010 & 0.0023 & 0.0066 \\
		& MoE-DeepPropNet& 0.0076 & 0.0031 & 0.0047 & 0.0099 & 0.0014 & 0.0065 & 0.0072 \\
		\hline
		
	\end{tabular}
\end{table}

\paragraph{}
From an overall perspective, both models demonstrate reliable predictive capability across the investigated parameter space. The S-DeepPropNet offers an efficient solution for individual property prediction, making it suitable for applications where only a limited number of plasma properties are required. In contrast, MoE-DeepPropNet provides a unified framework for simultaneous multi-properties evaluation, which is advantageous for coupled simulations and scenarios involving complex operating conditions. These characteristics allow flexible deployment depending on the specific requirements of plasma modeling.

\subsection{Predicting plasma properties of ternary gas mixtures}
\label{sec:sec3.2}
\paragraph{}
To further evaluate the performance of the proposed models for more complex plasma systems, the thermodynamic and transport properties of ternary C$_4$F$_7$N-CO$_2$-O$_2$ mixtures are considered. In this ternary system, the gas composition is defined by the mixing proportions of three species subject to the constraint that their sum equals 100\%. Accordingly, the input to the branch network is reduced to two independent variables, namely the mixing proportions of C$_4$F$_7$N (denoted as $C_a$) and CO$_2$ (denoted as $C_b$), while the O$_2$ proportion is determined as 100\% minus the sum of $C_a$ and $C_b$. The training dataset is constructed by varying both $C_a$ and $C_b$ from 10\% to 40\% with a uniform increment of 10\%, while the test set follows the same sampling strategy within the range of 15\% to 35\%. Both S-DeepPropNet and MoE-DeepPropNet are trained using logarithmically spaced pressures of 1.0, 1.58, 2.51, 3.98, 6.31, 10.0, and 15.85~bar. To account for the increased complexity of ternary plasma systems, S-DeepPropNet adopts six hidden layers with 200 neurons per layer, while MoE-DeepPropNet employs ten experts, each consisting of five hidden layers with 200 neurons. All other hyperparameter settings remain identical to those used in the binary case.

\begin{figure}[!b]
	\centering
	\includegraphics[width=15cm]{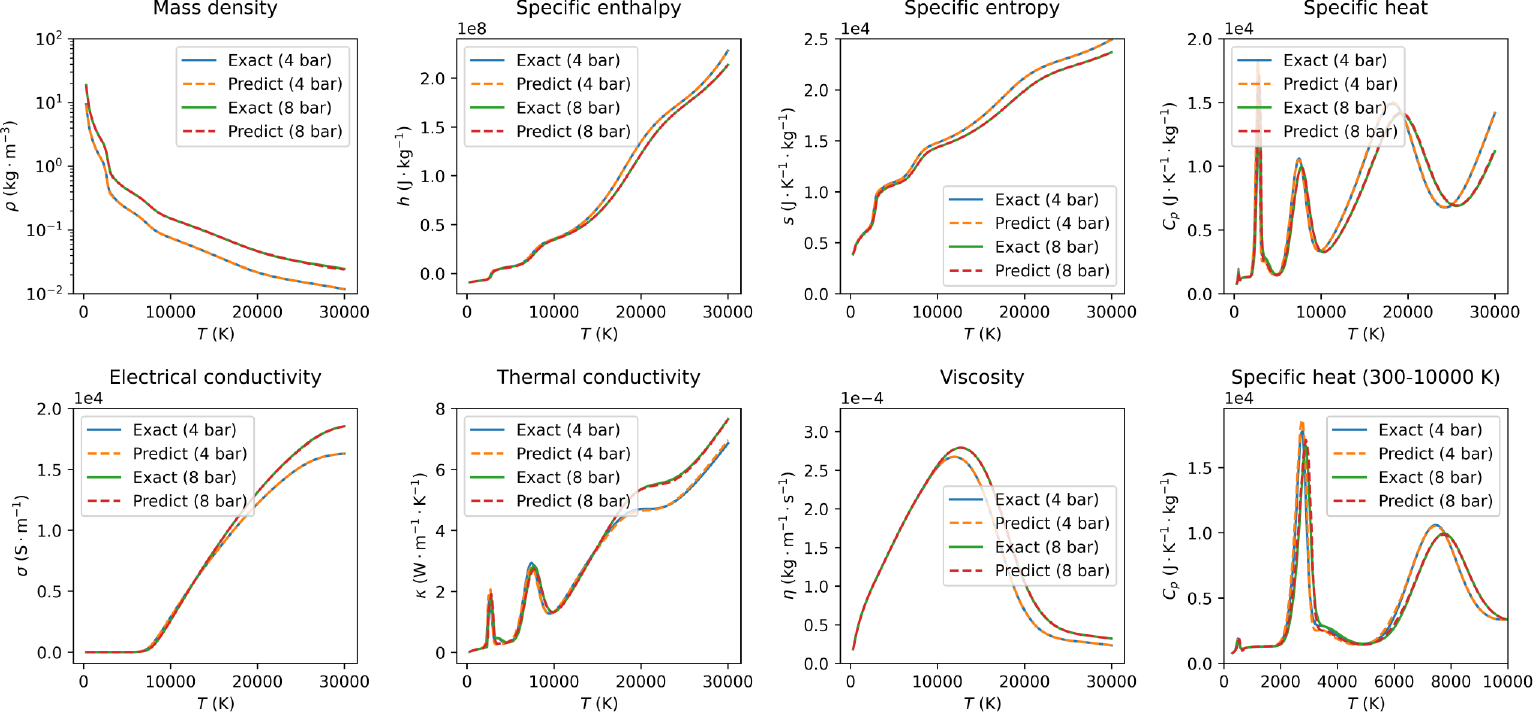}
	\caption{Prediction of thermodynamic and transport properties of C$_4$F$_7$N-CO$_2$-O$_2$ (25\%-25\%-50\%) plasmas at different pressures using MoE-DeepPropNet.}
	\label{fig:fig5}
\end{figure}

\paragraph{}
Fig.~\ref{fig:fig5} presents the predicted properties of an unseen C$_4$F$_7$N-CO$_2$-O$_2$ (25\%-25\%-50\%) plasma mixture at different pressures. Overall, the predicted results remain in close agreement with the reference data across the entire temperature range, indicating that the model retains its predictive capability in more complex ternary systems. Compared with the binary case, slightly larger deviations are observed near sharp localized peaks, particularly in the specific heat capacity $C_p$ and thermal conductivity $\kappa$. These regions correspond to strong thermochemical transitions associated with multi-species dissociation and ionization processes, which become more intricate in ternary mixtures due to the increased number of interacting species and reaction pathways~\cite{cite28}.

\paragraph{}
Despite these localized discrepancies, the overall thermodynamic and transport behaviors are well preserved. Since plasma properties are typically integrated over a wide temperature range in macroscopic simulations, such narrow-interval deviations are expected to have a limited impact on global flow and energy transport predictions. These results suggest that MoE-DeepPropNet remains suitable for practical applications involving complex multi-component plasmas.

\paragraph{}
Unlike binary systems, ternary mixtures are governed by two independently varying mixing ratios ($C_a$ and $C_b$), resulting in a higher-dimensional composition space. Since conventional two-dimensional property curves can only represent a limited subset of possible compositions, a tabular evaluation is adopted to quantify model performance across different mixing conditions. Specifically, the relative $L^2$ errors of seven thermodynamic and transport properties are averaged over eight unseen pressures (ranging from 2 to 16~bar with a 2~bar interval) for nine unseen gas mixing ratios. As summarized in Table \ref{tab:tab2}, the results indicate that MoE-DeepPropNet achieves consistent accuracy across a wide range of ternary compositions, with relative errors predominantly on the order of $10^{-3}$ to $10^{-2}$.

\begin{table}[htbp]
	\centering
	\caption{Prediction of thermodynamic and transport properties of C$_4$F$_7$N-CO$_2$-O$_2$ plasmas with different gas mixing ratios using MoE-DeepPropNet, averaged over eight pressures.}
	\label{tab:tab2}
	\begin{tabular}{|c|c|c|c|c|c|c|c|}
		\hline
		Mixture Ratio (C$_4$F$_7$N:CO$_2$:O$_2$) & $\rho$ & $h$ & $s$ & $C_p$ & $\sigma$ & $\kappa$ & $\eta$ \\
		\hline
		
		15:15:70 & 0.0226 & 0.0023 & 0.0023 & 0.0095 & 0.0010 & 0.018  & 0.0025 \\
		\hline
		15:25:60 & 0.0202 & 0.0024 & 0.0019 & 0.0098 & 0.0010 & 0.0235 & 0.0026 \\
		\hline
		15:35:50 & 0.0191 & 0.0026 & 0.0015 & 0.0138 & 0.0009 & 0.0282 & 0.0022 \\
		\hline
		25:15:60 & 0.0229 & 0.0047 & 0.0021 & 0.0142 & 0.0011 & 0.0257 & 0.0023 \\
		\hline
		25:25:50 & 0.0158 & 0.0049 & 0.0016 & 0.0127 & 0.0009 & 0.0219 & 0.0024 \\
		\hline
		25:35:40 & 0.0244 & 0.0050 & 0.0022 & 0.0200 & 0.0008 & 0.0195 & 0.0025 \\
		\hline
		35:15:50 & 0.0189 & 0.0031 & 0.0018 & 0.0389 & 0.0008 & 0.0222 & 0.0023 \\
		\hline
		35:25:40 & 0.0314 & 0.0026 & 0.0028 & 0.0380 & 0.0008 & 0.0204 & 0.0029 \\
		\hline
		35:35:30 & 0.0449 & 0.0029 & 0.0036 & 0.0368 & 0.0008 & 0.0202 & 0.0028 \\
		\hline
		
	\end{tabular}
\end{table}

\paragraph{}
Finally, a quantitative comparison between the S-DeepPropNet and the MoE-DeepPropNet is presented in Table \ref{tab:tab3}. The reported errors at unseen test pressures (4, 8, 12, and 16~bar) are averaged over all test gas mixing ratios. Both models achieve satisfactory predictive accuracy across the evaluated conditions. At moderate pressures (4-8~bar), the two models exhibit comparable performance for most properties, with only minor differences in error levels. More noticeable differences arise at higher pressures, where stronger nonlinear effects are present. In this regime, MoE-DeepPropNet generally provides improved accuracy, particularly for mass density $\rho$ (reduced from 0.0260 to 0.0093 in the overall average) and thermal conductivity $\kappa$ (from 0.0213 to 0.0113). From the overall average metrics, MoE-DeepPropNet achieves comparable or lower errors than S-DeepPropNet for most properties, although slight increases are observed for certain quantities, such as the specific heat capacity $C_p$.

\begin{table}[htbp]
	\centering
	\caption{Relative $L^2$ errors of thermodynamic and transport properties for C$_4$F$_7$N-CO$_2$-O$_2$ plasmas predicted by S-DeepPropNet and MoE-DeepPropNet at different pressures, averaged over test gas mixing ratios.}
	\label{tab:tab3}
	\begin{tabular}{|c|c|c|c|c|c|c|c|c|}
		\hline
		Pressure (bar) & Model & $\rho$ & $h$ & $s$ & $C_p$ & $\sigma$ & $\kappa$ & $\eta$ \\
		\hline
		
		\multirow{2}{*}{4}
		& S-DeepPropNet  & 0.0081 & 0.0013 & 0.0022 & 0.0277 & 0.0005 & 0.0113 & 0.0013 \\
		& MoE-DeepPropNet& 0.0101 & 0.0013 & 0.0021 & 0.0275 & 0.0014 & 0.0110 & 0.0014 \\
		\hline
		
		\multirow{2}{*}{8}
		& S-DeepPropNet  & 0.0082 & 0.0012 & 0.0021 & 0.0269 & 0.0006 & 0.0104 & 0.0011 \\
		& MoE-DeepPropNet& 0.0115 & 0.0018 & 0.0025 & 0.0273 & 0.0054 & 0.0103 & 0.0021 \\
		\hline
		
		\multirow{2}{*}{12}
		& S-DeepPropNet  & 0.0122 & 0.0021 & 0.0026 & 0.0269 & 0.0056 & 0.0100 & 0.0024 \\
		& MoE-DeepPropNet& 0.0086 & 0.0013 & 0.0021 & 0.0258 & 0.0006 & 0.0093 & 0.0010 \\
		\hline
		
		\multirow{2}{*}{16}
		& S-DeepPropNet  & 0.0088 & 0.0047 & 0.0039 & 0.0337 & 0.0041 & 0.0155 & 0.0058 \\
		& MoE-DeepPropNet& 0.0084 & 0.0012 & 0.0022 & 0.0299 & 0.0005 & 0.0142 & 0.0012 \\
		\hline
		
		\multirow{2}{*}{Overall}
		& S-DeepPropNet  & 0.0260 & 0.0039 & 0.0024 & 0.0209 & 0.0013 & 0.0213 & 0.0028 \\
		& MoE-DeepPropNet& 0.0093 & 0.0019 & 0.0024 & 0.0259 & 0.0024 & 0.0113 & 0.0022 \\
		\hline
		
	\end{tabular}
\end{table}

\paragraph{}
These results indicate that both models maintain reliable predictive capability, while the differences between the two architectures become more evident under increasingly complex operating conditions. This behavior can be attributed to the ability of the MoE architecture to distribute different input regimes across specialized experts, thereby reducing interference among simultaneously learned properties. In addition, since MoE-DeepPropNet evaluates all seven properties within a single forward pass, it avoids the need to train and deploy multiple independent models, resulting in improved computational efficiency. Overall, the proposed DeepPropNet framework provides a flexible and efficient solution for both single-property and multi-properties plasma modeling across a wide range of conditions.

\subsection{Thermal plasma simulation by combining finite volume method (FVM) and DeepPropNet}
\label{sec:sec3.3}
\begin{figure}[!t]
	\centering
	\includegraphics[width=10cm]{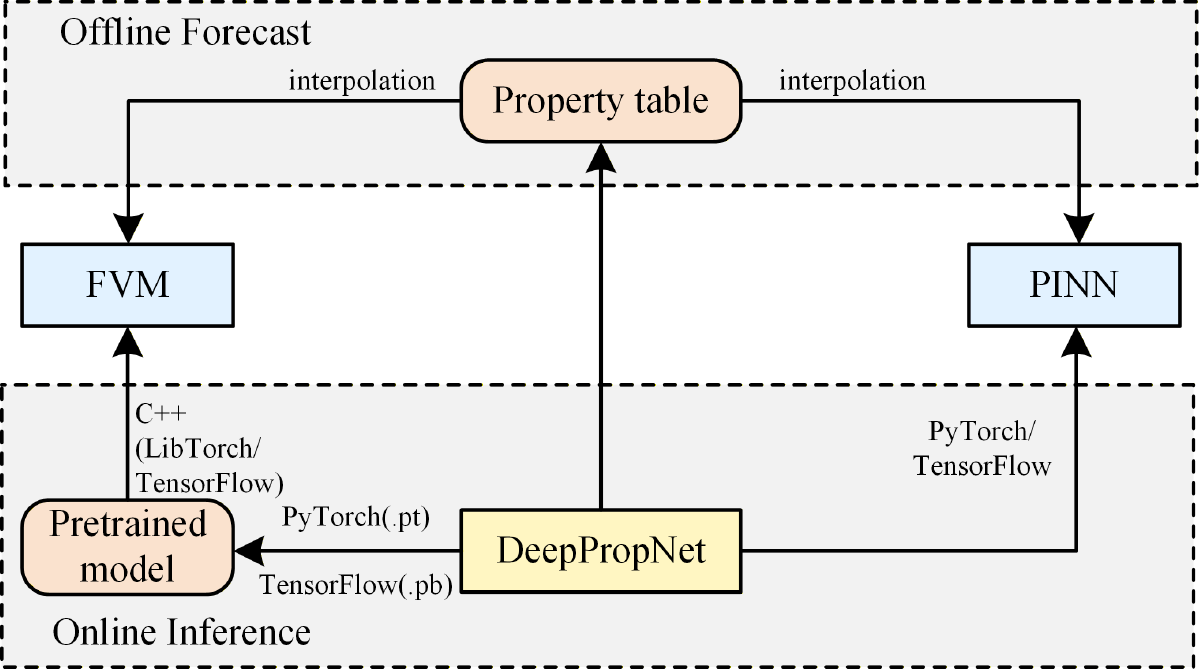}
	\caption{The integration of DeepPropNet with FVM and PINN frameworks, including the offline forecast and online inference strategies.}
	\label{fig:fig6}
\end{figure}

\paragraph{}
As illustrated in Fig.~\ref{fig:fig6}, the integration of DeepPropNet with plasma solvers can be implemented through two alternative strategies, namely offline forecast and online inference, where the implementation of the online inference strategy depends on the computational environment of the solver. For FVM-based solvers, the offline forecast strategy consists of generating plasma properties in advance using DeepPropNet and storing them as tabulated data, which are subsequently accessed through interpolation during the simulation. Alternatively, the online inference strategy can be realized through two implementation routes depending on the deep learning framework. One approach is to export the pretrained model as a ``.pb'' file and executed via the TensorFlow C++ interface. The other approach is to export the pretrained model as a ``.pt'' file and deployed in the C++ solver using the LibTorch interface for direct runtime inference. In the present section, the offline forecast strategy is adopted for coupling DeepPropNet with the FVM solver.

\begin{figure}
	\centering
	\includegraphics[width=10cm]{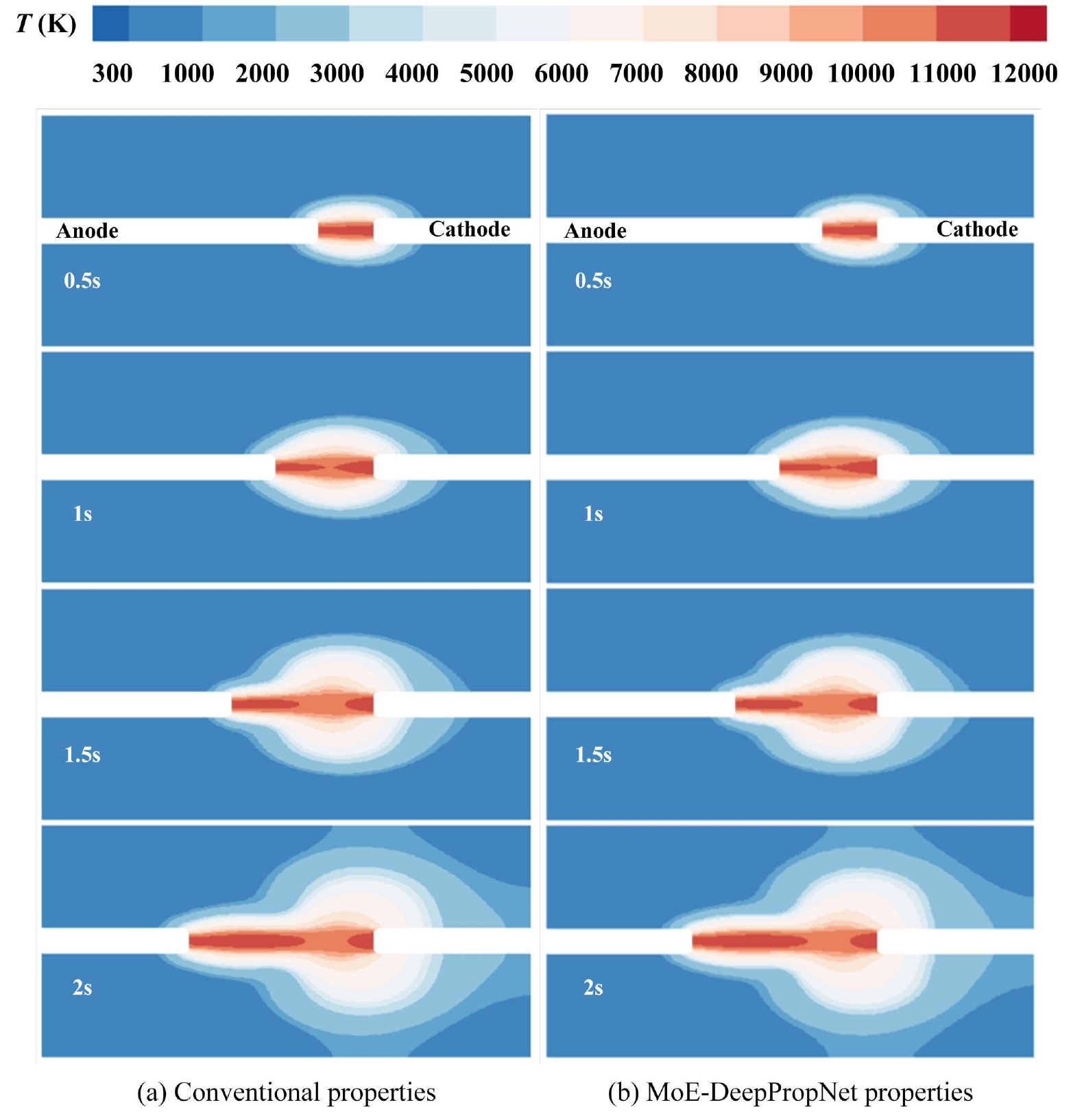}
	\caption{Comparison of temperature distributions obtained using plasma properties from traditional calculations and those predicted by MoE-DeepPropNet for a C$_4$F$_7$N-CO$_2$-O$_2$ (25\%-25\%-50\%) plasma mixture at 1 bar during the arc interruption process.}
	\label{fig:fig7}
\end{figure}

\paragraph{}
In this study, the thermal plasma (i.e., arc discharge) is primarily confined to the contact gap region. The computational domain is simplified to retain the essential physical characteristics of the arc while improving numerical efficiency. The detailed geometric configuration and solver settings are consistent with our previous work~\cite{cite29} and are not repeated here for brevity. Thermodynamic and transport properties in conventional FVM-based plasma simulations are provided in tabulated form and accessed through interpolation, and the MoE-DeepPropNet predictions are incorporated using the same procedure. The simulations are performed for a C$_4$F$_7$N-CO$_2$-O$_2$ (25\%-25\%-50\%) plasma mixture at an operating pressure of 1~bar. Fig.~\ref{fig:fig7} presents a qualitative comparison of the temperature distributions obtained using plasma properties derived from conventional calculations and those predicted by MoE-DeepPropNet during the arc interruption process. The two sets of results exhibit highly consistent spatial structures throughout the evolution of the arc. The arc core region, temperature gradients, and overall thermal expansion behavior are consistently reproduced, indicating that the properties generated by DeepPropNet preserve the essential thermophysical characteristics required for plasma fluid simulations.

\begin{figure}
	\centering
	\includegraphics[width=12cm]{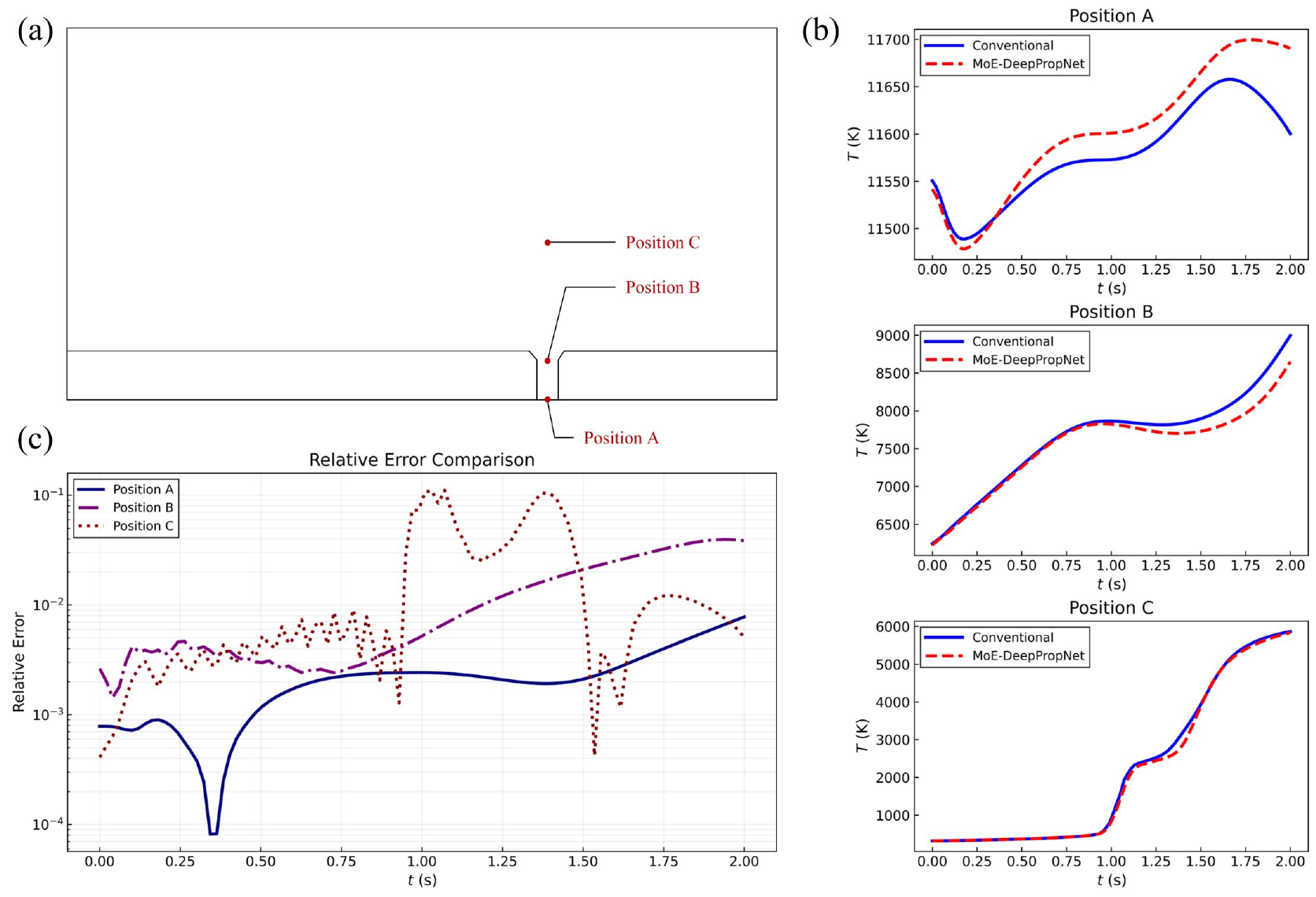}
	\caption{(a) Computational domain and monitoring locations. (b) Temperature evolution at Position A (arc center), Position B (near-axis region), and Position C (arc-edge region). (c) Relative error of temperature at the three locations. Results are obtained for a C$_4$F$_7$N-CO$_2$-O$_2$ (25\%-25\%-50\%) plasma mixture at 1 bar using plasma properties from conventional calculations and MoE-DeepPropNet.}
	\label{fig:fig8}
\end{figure}

\paragraph{}
To quantify the agreement between the two cases, the temporal evolution of temperature at representative locations within the computational domain is analyzed. As shown in Fig.~\ref{fig:fig8}, three monitoring points are selected to characterize different regions of the arc, including the arc center (Position A), the near-axis region (Position B), and the arc-edge region (Position C). The results demonstrate that the temperature profiles obtained using MoE-DeepPropNet generated properties are in close agreement with those based on conventionally calculated properties over the entire simulation period. The agreement is particularly strong in the arc core region, where accurate representation of ionization and energy transport processes is critical. Minor deviations can be observed at specific time instances and in peripheral regions. These discrepancies mainly originate from the residual prediction errors in the MoE-DeepPropNet generated plasma properties, especially in regions with strong nonlinear property variations, and their subsequent propagation through the numerical simulation. However, these discrepancies remain limited and do not affect the overall physical evolution of the arc plasma.

\paragraph{}
Overall, the results confirm that the plasma properties generated by MoE-DeepPropNet can reliably reproduce both the spatial distribution and temporal evolution of arc plasma behavior within the FVM framework, demonstrating sufficient fidelity for macroscopic plasma simulations. Compared with conventional approaches based on precomputed datasets derived from physical models, MoE-DeepPropNet provides a data-driven representation of plasma properties that remains fully compatible with interpolation-based implementations. This enables the evaluation of plasma properties over a continuous parameter space without repeated evaluations of the underlying physical models. The present coupling strategy therefore offers a practical pathway for integrating operator learning-based surrogate models into large-scale plasma simulations.

\subsection{Thermal plasma simulation by combining physics-informed neural networks (PINNs) and DeepPropNet}
\label{sec:sec3.4}
\paragraph{}
In contrast to the FVM solver, physics-informed neural networks (PINNs) are usually natively implemented in deep learning frameworks (e.g., PyTorch), which enables direct integration of DeepPropNet within the same computational framework. In this setting, MoE-DeepPropNet is embedded into the PINN model following the online inference strategy illustrated in Fig.~\ref{fig:fig6}, allowing plasma properties to be evaluated in real time during the forward pass without relying on precomputed data. Following our previous work, the governing equations of thermal plasmas are embedded into the loss function of neural networks, enabling the solution of partial differential equations in a mesh-free manner. A detailed description of the PINN framework can be found in our literature~\cite{cite30,cite31}.

\paragraph{}
Using the framework of coefficient-subnet PINN (CS-PINN)~\cite{cite30}, the thermodynamic and transport properties appearing in the governing equations are treated as state-dependent coefficients $\lambda$, which are functions of temperature, pressure, and gas composition. The general form of the plasma model can be expressed as:

\begin{equation}
\label{equ:equ8}
f\left(t, x, \frac{\partial u}{\partial t}, \frac{\partial u}{\partial x}, \cdots, \lambda \right) = 0
\end{equation}

where $x$ denotes the spatial coordinates and $t$ represents time, $u$ denotes the solution variables, and $\lambda$ represents the plasma properties. In conventional implementations, these coefficients are typically obtained from tabulated datasets and either evaluated through interpolation or approximated by separate pre-trained surrogate models for individual properties. In the present work, MoE-DeepPropNet is incorporated as a unified model to provide $\lambda$ directly as continuous functions of the local thermodynamic state, thereby eliminating the need for multiple independent property models while maintaining consistency within the PINN framework.

\paragraph{}
To evaluate the performance of the proposed approach, a one-dimensional arc plasma model is first considered. In this case, the steady-state arc described by the Elenbaas-Heller equation is solved, which governs the radial temperature distribution in a cylindrical arc column. The equation can be written as:

\begin{equation}
\label{equ:equ9}
\frac{1}{r}\frac{\partial}{\partial r}\left( r \kappa \frac{\partial T}{\partial r} \right) + \sigma \frac{I^2}{g^2} - E_{\mathrm{rad}} = 0
\end{equation}

\begin{equation}
\label{equ:equ10}
g = \int_{0}^{R} 2\pi r \sigma \, dr
\end{equation}

where $r$ is the radial coordinate, $T$ is the arc temperature, $\sigma$ is the electrical conductivity, $\kappa$ is the thermal conductivity, $I$ is the arc current, $g$ is the arc conductance, and $E_{\mathrm{rad}}$ represents the radiative energy loss. The plasma properties $\sigma$ and $\kappa$ are provided by the MoE-DeepPropNet model.

\begin{figure}
	\centering
	\includegraphics[width=12cm]{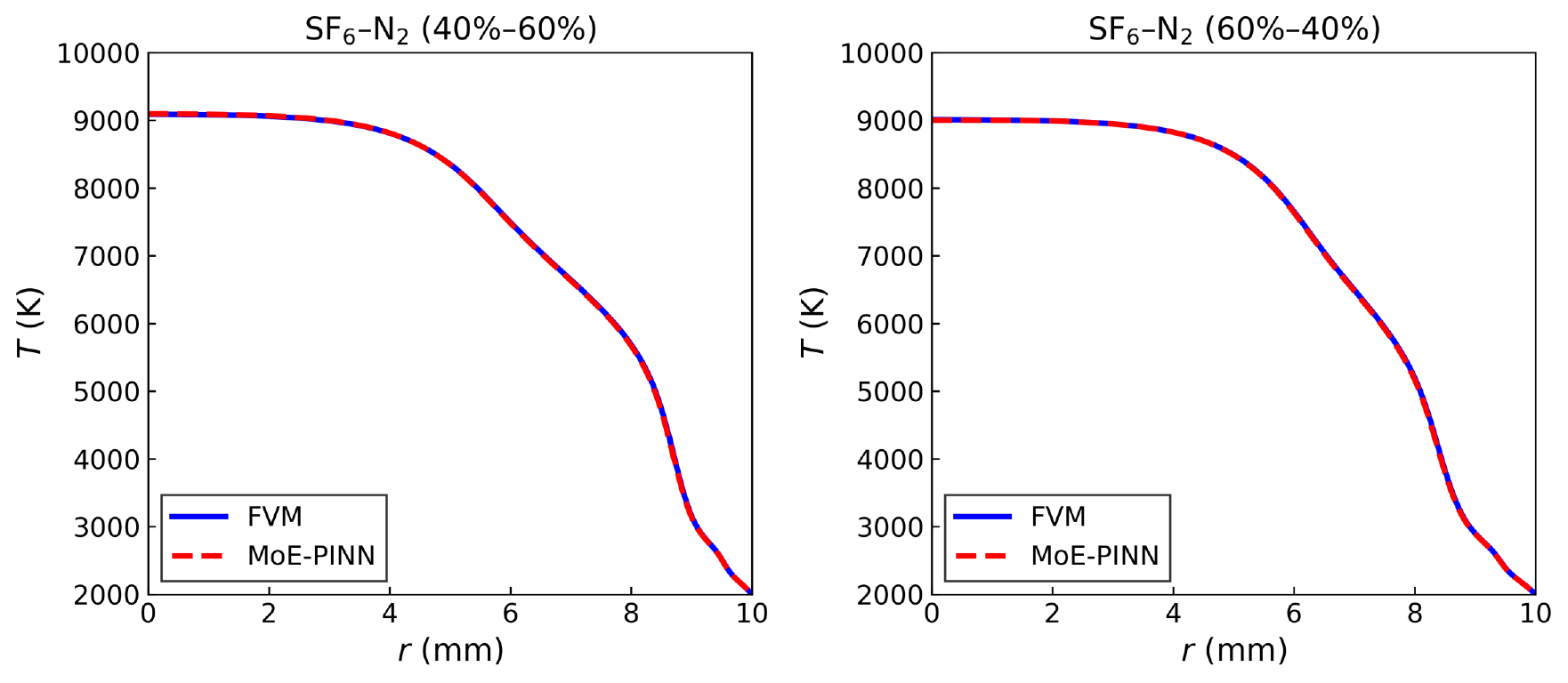}
	\caption{Prediction of radial temperature distribution of a one-dimensional stationary arc for SF$_6$-N$_2$ mixtures at 1~bar using the MoE-PINN, with comparison to reference FVM solutions for two representative compositions (40\%-60\% and 60\%-40\%). The relative $L^2$ errors are on the order of $10^{-3}$.}
	\label{fig:fig9}
\end{figure}

\paragraph{}
Fig.~\ref{fig:fig9} shows the radial temperature distributions for SF$_6$-N$_2$ mixtures at 1~bar with two representative compositions, namely 40\%-60\% and 60\%-40\%. The results obtained from the PINN model coupled with MoE-DeepPropNet (MoE-PINN) are compared with those from the reference finite volume method (FVM) solution. The MoE-PINN employs a feed-forward neural network with six hidden layers and 50 neurons per layer. It can be observed that the predicted temperature profiles are in close agreement with the reference results over the entire radial domain. The arc core region, as well as the steep temperature gradients near the arc boundary, are accurately captured. The relative $L^2$ errors are on the order of $10^{-3}$.

\paragraph{}
For the transient case, a one-dimensional arc model without radial velocity is considered. In this case, the governing equation describes the temporal evolution of the arc temperature and can be written as:

\begin{equation}
\label{equ:equ11}
\rho C_p \frac{\partial T}{\partial t}
=
\frac{1}{r}\frac{\partial}{\partial r}\left( r \kappa \frac{\partial T}{\partial r} \right)
+ \sigma \frac{I^2}{g^2}
- E_{\mathrm{rad}}
\end{equation}

where $\rho$ is the mass density and $C_p$ is the specific heat at constant pressure, and the other variables have the same definitions as in Eq.~(9). The plasma properties ($\rho$, $C_p$, $\sigma$, $\kappa$) are provided by the MoE-DeepPropNet model.

\begin{figure}[!b]
	\centering
	\includegraphics[width=12cm]{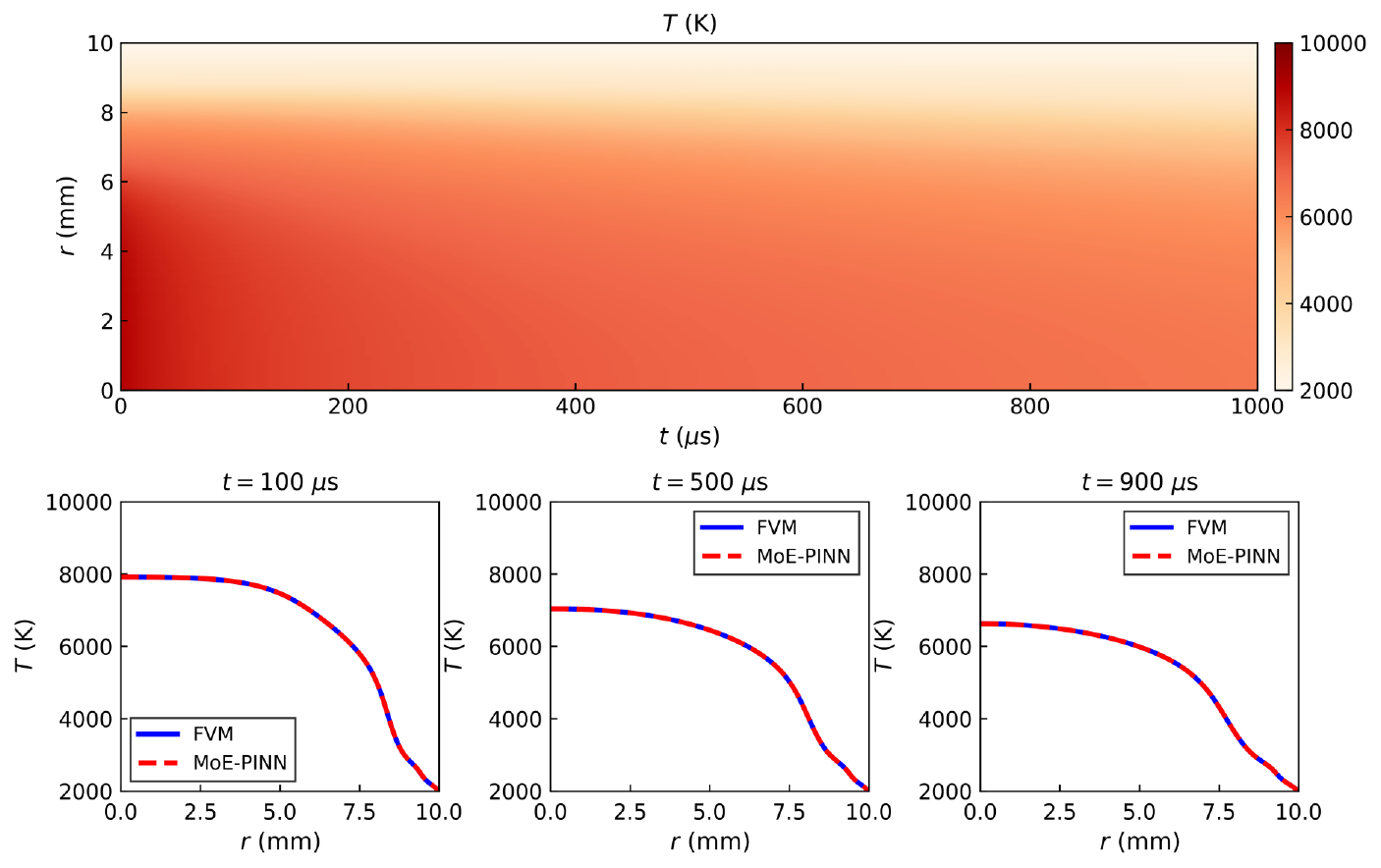}
	\caption{Transient arc simulation for an SF$_6$-N$_2$ (60\%-40\%) plasma mixture at 1~bar. The spatiotemporal evolution of temperature is predicted by the MoE-PINN, together with radial temperature distributions at selected time instants (t = 100 $\mu$s, 500 $\mu$s, and 900 $\mu$s) compared with reference FVM solutions. The relative $L^2$ errors at representative time instants are on the order of $10^{-4}$.}
	\label{fig:fig10}
\end{figure}

\paragraph{}
Fig.~\ref{fig:fig10} presents the transient arc simulation for an SF$_6$-N$_2$ (60\%-40\%) plasma mixture at 1~bar. The MoE-PINN employs a feed-forward neural network with six hidden layers and 200 neurons per layer. The spatiotemporal evolution of temperature shows a gradual cooling process, with the temperature in the arc core decreasing over time while maintaining a smooth spatial distribution. This behavior is physically consistent with the expected evolution of a decaying arc. To further assess the accuracy of the proposed framework, radial temperature distributions at selected time instants are compared with reference solutions. The results demonstrate that the MoE-PINN predictions are in close agreement with the reference solutions at all considered time instants. The relative $L^2$ errors at representative time instants are on the order of $10^{-4}$.

\paragraph{}
To further quantify the convergence behavior of the PINN models with different plasma property representations, Table \ref{tab:tab4} summarizes the number of training epochs required to reach specified loss thresholds. For the steady-state case, the MoE-PINN demonstrates a significantly accelerated convergence compared with the lookup-based PINN. In particular, the number of epochs required to reach the $10^{-1}$ and $10^{-2}$ loss levels is reduced by nearly an order of magnitude. Moreover, while the lookup-based PINN fails to reach the $10^{-3}$ threshold within the prescribed training budget, the MoE-PINN successfully attains this level. For the transient case, both approaches exhibit similar convergence behavior during the initial training stage, as evidenced by comparable epochs required to reach the $10^{-1}$ and $10^{-2}$ levels. However, clear differences emerge at lower loss levels. The MoE-PINN not only reaches the $10^{-4}$ threshold with fewer training epochs, but also achieves the $5\times10^{-5}$ level, which is unattainable for the lookup-based counterpart within the same training budget.

\paragraph{}
The improved convergence behavior can be attributed to the continuous and differentiable representation of plasma properties provided by MoE-DeepPropNet, which facilitates smoother gradient propagation during PINN training compared with interpolation-based lookup tables. This results in more stable training dynamics and improved solution accuracy. Overall, the successful implementation of both the offline forecast and online inference strategies demonstrates the flexibility of DeepPropNet for integration with different simulation frameworks. The model can either provide precomputed property data for interpolation-based implementations or enable direct neural network inference. The latter can be deployed in C++-based solvers via different interfaces depending on the deep learning framework, for example, using the LibTorch or the TensorFlow C++ interface, or integrated natively within PyTorch or TensorFlow-based PINN frameworks. This flexibility provides a practical pathway for incorporating operator learning-based surrogate models into practical plasma simulations.

\begin{table}[htbp]
	\centering
	\caption{Number of training epochs required to reach specified loss thresholds for SF$_6$-N$_2$ (60\%-40\%) plasma mixture at 1~bar: Comparison between lookup-based PINN and MoE-PINN.}
	\label{tab:tab4}
	\begin{tabular}{|c|c|c|c|c|c|c|}
		\hline
		Case & Method & $10^{-1}$ & $10^{-2}$ & $10^{-3}$ & $10^{-4}$ & $5\times10^{-5}$ \\
		\hline
		
		\multirow{2}{*}{Steady}
		& Lookup-based PINN & 67565 & 89532 & / & / & / \\
		& MoE-PINN          &  8388 & 65877 & 92749 & / & / \\
		\hline
		
		\multirow{2}{*}{Transient}
		& Lookup-based PINN &   143 &  4315 & 13434 & 39817 & / \\
		& MoE-PINN          &   157 & 12726 & 14517 & 34597 & 35386 \\
		\hline
		
	\end{tabular}
\end{table}

\section{Conclusions}
\label{sec:sec4}
\paragraph{}
In this work, an operator learning-based framework, namely DeepPropNet, is proposed for fast and accurate prediction of thermodynamic and transport properties of thermal plasmas. Two architectures are developed, including the single-property model S-DeepPropNet and the multi-properties model MoE-DeepPropNet. The main conclusions are summarized as follows:

\paragraph{}
1.\ The proposed models can accurately capture the complex nonlinear dependence of plasma properties on temperature, pressure, and gas composition. For both binary (SF$_6$-N$_2$) and ternary (C$_4$F$_7$N-CO$_2$-O$_2$) mixtures, the predicted results show excellent agreement with reference data, with relative $L^2$ errors predominantly on the order of $10^{-3}$--$10^{-2}$. The S-DeepPropNet is well suited for single-property prediction tasks with lower computational complexity, while the MoE-DeepPropNet improves robustness under complex conditions and enables efficient multi-property prediction within a single forward pass.

\paragraph{}
2.\ The effectiveness of DeepPropNet is attributed to its ability to learn a continuous mapping from plasma operating conditions to physical properties, providing a smooth and differentiable representation across the parameter space. The Mixture-of-Experts architecture further mitigates interference among multiple prediction tasks by dynamically routing inputs to specialized subnetworks, leading to improved stability and accuracy under strongly nonlinear conditions.

\paragraph{}
3.\ A unified integration framework is established to couple DeepPropNet with both finite volume method (FVM) solvers and physics-informed neural networks (PINNs) through offline forecast and online inference strategies. In the FVM framework, DeepPropNet-generated properties are incorporated in tabulated form and accessed through interpolation, reproducing the spatial distribution and temporal evolution of arc behavior with high fidelity. In the PINN framework, DeepPropNet is directly embedded into the PINN model, enabling continuous and differentiable property evaluation during training and inference. This integration strategy allows the model to be flexibly deployed across different computational environments, including interpolation-based tabulated implementations, direct inference in C++-based solvers via deep learning interfaces, and native integration within PyTorch or TensorFlow-based PINN frameworks.

\paragraph{}
Overall, the proposed framework provides an effective approach for incorporating operator learning-based surrogate models into plasma simulations. Future work will focus on extending the framework to more complex plasma conditions, including non-equilibrium effects and a wider range of gas mixtures, as well as exploring its application in higher-dimensional and fully coupled plasma simulation problems.

\section*{Data availability}
\label{sec:data availability}
\paragraph{}
The data that support the findings of this study are available from the corresponding author upon reasonable request.

\section*{Acknowledgments}
\label{sec:acknowledgments}
\paragraph{}
This work was supported in part by the National Natural Science Foundation of China (92470102) and the Natural Science Foundation of Jiangsu Province (BK20231427).

\bibliographystyle{unsrt}


\end{document}